\documentclass[12pt]{article}

\usepackage{amsmath,amssymb}
\usepackage{amsfonts}
\usepackage{graphicx}
\usepackage{a4wide}

\usepackage[T2A]{fontenc}
\usepackage[cp1251]{inputenc}
\usepackage[russian]{babel}

\begin{document}

\begin{center} {\bf M. V. Burnashev} \end{center}

\vskip 0.4cm

\begin{center}
{\large\bf ON THE DISTRIBUTION OF THE STATISTICAL SUM RELATED TO BSC}
\footnote[1]{The research is supported by the MSHE project No. FSMG-2024-0048}.
\end{center}

{\begin{quotation} \normalsize For the information transmission a
binary symmetric channel with noiseless feedback is used. The transmission of
exponential number of messages is \\
considered. The best possible decoding
error probability exponent (reliability function) is found.

\end{quotation}}

\vskip 0.7cm

\begin{center}
{\bf \S\,1. Introduction and main results}
\end{center}


Assume that $\boldsymbol{x}_{j}= (x_{j}(1),\ldots,x_{j}(n))$, $j=1,\ldots, M$
are i.i.d. random equiprobable binary $n$-vectors with
$p(x_{j}(l) = 0) = p(x_{j}(l) =1) = 1/2$, $l=1,\ldots,n$, $j=1,\ldots, M$.
Denote by $w(\boldsymbol{x}_{j})$, $j=1,\ldots,M$ - weights (i.e. the number of
ones) of binary $n$-vectors $\{\boldsymbol{x}_{j}\}$. Consider the random sum
\begin{equation} \label{sum2}
\begin{gathered}
S(z,M,n) = \sum\limits_{j=1}^{M}z^{w_{j}(n)}, \qquad z > 0.
\end{gathered}
\end{equation}
Although all results below are non-asymptotic in $n,M$, they are mostly oriented
to the case $n \to \infty$ and $M=e^{Rn}$, $R > 0$.

The main results of the note constitute Theorem 1 and Theorem 2 for the cases
$z \leq 1$ and $z \geq 1$, respectively (to simplify formulas we do not use
integer parts signs).

{\bf Theorem 1}. 1) For $0 \leq z \leq 1$ and $z^{n/2} \leq A \leq 1$
the following bounds hold
\begin{equation} \label{Theo1}
\begin{gathered}
- \frac{\ln(n+1)}{n} \leq \frac{1}{Mn}\ln{\mathbf P}\{S(z,M,n)\geq MA\} + \ln 2 -
h(a_{0})\leq \frac{\ln n}{n}, \\
a_{0} = \frac{\ln A}{n\ln z}, \qquad h(x) = -x\ln x -(1-x)\ln(1-x), \quad
0 \leq a_{0} \leq 1/2.
\end{gathered}
\end{equation}
2) For $z^{n} \leq A \leq z^{n/2}$ the following bound holds
\begin{equation} \label{Theo1a}
\begin{gathered}
\frac{1}{Mn}\ln{\mathbf P}\{S(z,M,n) \leq MA\} \leq h(a_{0})-\ln 2 +
\frac{\ln(n+1)}{n}, \quad 1/2 \leq a_{0} \leq 1.
\end{gathered}
\end{equation}

{\bf Theorem 2}. 1) For $z \geq 1$ and $z^{n/2} \leq A \leq z^{n}$
the following bounds hold
\begin{equation} \label{Theo2}
\begin{gathered}
- \frac{\ln(n+1)}{n} \leq \frac{1}{Mn}\ln{\mathbf P}\{S(z,M,n)\geq MA\} + \ln 2 -
h(a_{1})\leq \frac{\ln n}{n}, \\
a_{1} = 1-\frac{\ln A}{n\ln z}, \quad 0 \leq a_{1} \leq 1/2.
\end{gathered}
\end{equation}

2) For $z \geq 1$ and $1 \leq A \leq z^{n/2}$ the following bound holds
\begin{equation} \label{Theo2a}
\begin{gathered}
\frac{1}{Mn}\ln{\mathbf P}\{S(z,M,n) \leq MA\} \leq h(a_{1})-\ln 2 +
\frac{\ln(n+1)}{n}, \quad 1/2 \leq a_{1} \leq 1.
\end{gathered}
\end{equation}

In particular, we get from \eqref{Theo1}-\eqref{Theo1a} and
\eqref{Theo2}-\eqref{Theo2a}

{\bf Corollary 1}. For any $z > 0$ the following inequality holds
\begin{equation} \label{Cor1a}
\begin{gathered}
{\mathbf P}\left\{\left|\ln\frac{S(z,M,n)}{Mz^{n/2}}\right| \geq
\sqrt{n\ln(n+1)}|\ln z|\right\} \leq (n+1)^{-M}.
\end{gathered}
\end{equation}

{\bf Remark}. It follows from \eqref{Cor1a} that if $M \sim e^{Rn}$, $R > 0$,
then $S(z,M,n) \sim Mz^{n/2}$ for any $z > 0$ with very high probability.

\begin{center}
{\bf \S\,2. Proofs}
\end{center}

{\bf 1. Proof of Theorem 1}. Note that
\begin{equation} \label{sum3}
\begin{gathered}
{\mathbf E}S(z,M,n) = M {\mathbf E}z^{w_{1}(n)}= M2^{-n}(1+z)^{n}, \\
{\mathbf E}e^{\lambda S(z,M,n)} = \left({\mathbf E}e^{\lambda z^{w_{1}(n)}}
\right)^{M}, \qquad {\mathbf E}e^{\lambda z^{w_{1}(n)}} =
2^{-n}\sum_{l=0}^{n}\binom{n}{l}e^{\lambda z^{l}}.
\end{gathered}
\end{equation}

Below, for binomial coefficients $\binom{n}{k}$ we use known lower and upper
bounds \cite[formula (12.40)]{CT}
\begin{equation} \label{binom1}
\begin{gathered}
\frac{1}{(n+1)}e^{nh(k/n)} \leq \binom{n}{k} \leq e^{nh(k/n)},
\quad 0\leq k \leq n, \\
h(x) = -x\ln x -(1-x)\ln(1-x).
\end{gathered}
\end{equation}

By Chebyshev exponential inequality (Chernov bound) and the last relation of
\eqref{sum3} we have for any $A \geq 0$
\begin{equation} \label{Cheb}
\begin{gathered}
{\mathbf P}\{S(z,M,n) \geq MA\} \leq \min_{\lambda \geq 0}
\left\{e^{-\lambda MA}{\mathbf E}e^{\lambda S(z,M,n)}\right\} = \\
= 2^{-Mn}\min_{\lambda \geq 0}\left\{e^{-\lambda MA}
\left[\sum_{l=0}^{n}\binom{n}{l}e^{\lambda z^{l}}\right]^{M}\right\},
\end{gathered}
\end{equation}
and then by \eqref{binom1} with $l=an$, $0 \leq a \leq 1$
\begin{equation} \label{Cheb1}
\begin{gathered}
\frac{1}{M}\ln{\mathbf P}\{S(z,M,n) \geq MA\} \leq -n\ln 2 +
\min_{\lambda \geq 0}\left\{-\lambda A+
\ln \left[\sum_{l=0}^{n}\binom{n}{l}e^{\lambda z^{l}}\right]\right\} \leq \\
\leq -n\ln 2 +\ln(n+1) + \min_{\lambda \geq 0}\max_{0\leq l\leq n}
\left\{-\lambda A +\ln\left[\binom{n}{l}e^{\lambda z^{l}}\right]\right\}
\leq \\
\leq -n\ln 2 +\ln(n+1) + \min_{\lambda \geq 0}\max_{0\leq a \leq 1}
f(\lambda,a,z),
\end{gathered}
\end{equation}
where we denoted
\begin{equation} \label{Cheb11}
\begin{gathered}
f(\lambda,a,z) = -\lambda A + nh(a) + \lambda z^{an}.
\end{gathered}
\end{equation}
Note that
\begin{equation} \label{Cheb12}
\begin{gathered}
f'_{a}(\lambda,a,z) = n\left[\ln\frac{1-a}{a} + \lambda z^{an}\ln z\right],
\quad f''_{aa}(\lambda,a,z) =
n\left[n\lambda z^{an}\ln^{2}z-\frac{1}{a(1-a)}\right].
\end{gathered}
\end{equation}
Denote by $a_{0}=a_{0}(\lambda)$ a root of the equation $f'_{a}(\lambda,a,z) =0$,
i.e.
\begin{equation} \label{Cheb1a}
\begin{gathered}
\ln\frac{1-a_{0}}{a_{0}} + \lambda z^{a_{0}n}\ln z = 0.
\end{gathered}
\end{equation}
Since $f'_{a}(\lambda,a,z) < 0$ for $a\geq 1/2$, $\lambda > 0$,
we may assume that $a \leq 1/2$ in \eqref{Cheb1}.
Then $a_{0}(0)=1/2$ and $a_{0}(\lambda) < 1/2$ for $\lambda > 0$.
We have by \eqref{Cheb1a}
\begin{equation} \label{Cheb2aa}
\begin{gathered}
\frac{d}{d\lambda}f(\lambda,a_{0}(\lambda),z) =
- A + z^{a_{0}n} + n\left[\ln\frac{1-a_{0}}{a_{0}} + \lambda z^{a_{0}n}\ln z
\right](a_{0})'_{\lambda} = - A + z^{a_{0}(\lambda)n}.
\end{gathered}
\end{equation}
By \eqref{Cheb2aa}, we set $\lambda_{0} \geq 0$ such that
\begin{equation} \label{Cheb3a}
\begin{gathered}
z^{a_{0}(\lambda_{0})n} =  A, \qquad a_{0}(\lambda_{0}) = \frac{\ln A}{n\ln z}.
\end{gathered}
\end{equation}
Since we need $0 \leq a_{0}=a_{0}(\lambda) \leq 1/2$, then for \eqref{Cheb3a}
we need $z^{n/2} \leq A \leq 1$. Therefore, by \eqref{Cheb1}-\eqref{Cheb3a}
\begin{equation} \label{Cheb3b}
\begin{gathered}
\frac{1}{Mn}\ln{\mathbf P}\{S(z,M,n) \geq MA\} \leq h[a_{0}(\lambda_{0})] -
\ln 2+ \frac{\ln(n+1)}{n}, \quad z^{n/2} \leq A \leq 1.
\end{gathered}
\end{equation}

To get a similar to \eqref{Cheb3b} lower bound, we consider the case when all
binary $n$-vectors $\{\boldsymbol{x}_{j}\}$ in the sum \eqref{sum2} have weight
$l_{0}$ such that $A=z^{l_{0}}$. Then by \eqref{binom1} we have for any
$A \geq z^{n}$
\begin{equation} \label{low1}
\begin{gathered}
\frac{1}{Mn}\ln{\mathbf P}\{S(z,M,n) \geq MA\} \geq
\frac{1}{Mn}\ln\left[2^{-n}\binom{n}{l_{0}}\right]^{M} \geq \\
\geq h\left(\frac{\ln A}{n\ln z}\right) - \ln 2- \frac{\ln(n+1)}{n} =
h[a_{0}(\lambda_{0})] -\ln 2 - \frac{\ln(n+1)}{n}.
\end{gathered}
\end{equation}
From \eqref{Cheb3b} and \eqref{low1} we get \eqref{Theo1}.

In order to prove \eqref{Theo1a}, similarly to \eqref{Cheb} and \eqref{Cheb1}
we have for any $A \geq 0$
\begin{equation} \label{Cheb2}
\begin{gathered}
{\mathbf P}\{S(z,M,n) \leq MA\} \leq \min_{\lambda \geq 0}
\left\{e^{\lambda MA}{\mathbf E}e^{-\lambda S(z,M,n)}\right\} = \\
= 2^{-Mn}\min_{\lambda \geq 0}\left\{e^{\lambda MA}
\left[\sum_{l=0}^{n}\binom{n}{l}e^{-\lambda z^{l}}\right]^{M}\right\}.
\end{gathered}
\end{equation}
Then with $l=an$, $0 \leq a \leq 1$
\begin{equation} \label{Cheb2a}
\begin{gathered}
\frac{1}{M}\ln{\mathbf P}\{S(z,M,n) \leq MA\} \leq -n\ln 2 +
\min_{\lambda \geq 0}\left\{\lambda A +
\ln \left[\sum_{l=0}^{n}\binom{n}{l}e^{-\lambda z^{l}}\right]\right\} \leq \\
\leq -n\ln 2 +\ln(n+1) + \min_{\lambda \geq 0}\max_{0\leq l\leq n}
\left\{\lambda A +\ln\left[\binom{n}{l}e^{-\lambda z^{l}}\right]\right\}
\leq \\
\leq -n\ln 2 +\ln(n+1) + \min_{\lambda \geq 0}\max_{0\leq a \leq 1}g(\lambda,a),
\end{gathered}
\end{equation}
where we denoted
\begin{equation} \label{Cheb2c}
\begin{gathered}
g(\lambda,a) = \lambda A + nh(a) - \lambda z^{an}.
\end{gathered}
\end{equation}
Note that
\begin{equation} \label{Cheb2d}
\begin{gathered}
g'_{a}(\lambda,a) = n\left[\ln\frac{1-a}{a} - \lambda z^{an}\ln z\right].
\end{gathered}
\end{equation}
Denote by $a_{1}=a_{1}(\lambda)$ a root of the equation $g'_{a}(\lambda,a) =0$,
i.e.
\begin{equation} \label{Cheb2f}
\begin{gathered}
\ln\frac{1-a_{1}}{a_{1}} - \lambda z^{a_{1}n}\ln z = 0.
\end{gathered}
\end{equation}
Then $a_{1}(0)=1/2$ and $a_{1}(\lambda) > 1/2$, $\lambda > 0$.
We have by \eqref{Cheb2f}
\begin{equation} \label{Cheb2g}
\begin{gathered}
\frac{d}{d\lambda}g(\lambda,a_{1}(\lambda)) =
A_{1} - z^{a_{1}n} + n\left[\ln\frac{1-a_{1}}{a_{1}} - \lambda z^{a_{1}n}\ln z
\right](a_{1})'_{\lambda} = A_{1} - z^{a_{1}(\lambda)n}.
\end{gathered}
\end{equation}
By \eqref{Cheb2g}, we set $\lambda_{1} \geq 0$ such that
$z^{a_{1}(\lambda_{1})n} =  A$, i.e. (see also \eqref{Cheb3a})
\begin{equation} \label{Cheb2h}
\begin{gathered}
a_{1}(\lambda_{1}) = a_{0}(\lambda_{0}) =\frac{\ln A}{n\ln z} \geq \frac{1}{2}.
\end{gathered}
\end{equation}
Then we get from \eqref{Cheb2a} and \eqref{Cheb2h} for
 $z^{n} \leq A \leq z^{n/2}$
\begin{equation} \label{Cheb2k}
\begin{gathered}
\frac{1}{Mn}\ln{\mathbf P}\{S(z,M,n) \leq MA\}\leq h(a_{0}(\lambda_{0}))-\ln 2 +
\frac{\ln(n+1)}{n}.
\end{gathered}
\end{equation}

As a result, we get \eqref{Theo1a} from \eqref{Cheb2k}, what proves the Theorem.

{\bf 2. Proof of Theorem 2}. Note that for any $z > 0$ and any $j=1,\ldots,M$
random variables $z^{w_{j}(n)}$ and $z^{n-w_{j}(n)}$ have the same distribution
functions. Therefore, random sums $S(z,M,n)$ and $z^{n}S(1/z,M,n)$ have also
the same distribution functions. Then we have by relations \eqref{Theo1} for
$z \geq 1$ and $z^{n/2} \leq A \leq z^{n}$
\begin{equation} \label{Theo2aa}
\begin{gathered}
\frac{1}{Mn}\ln{\mathbf P}\{S(z,M,n)\geq MA\} =
\frac{1}{Mn}\ln{\mathbf P}\{S(1/z,M,n)\geq MAz^{-n}\} \leq \\
\leq h(a_{1}) - \ln 2 + \frac{\ln n}{n}, \\
a_{1} = -\frac{\ln(Az^{-n})}{n\ln z} = 1- \frac{\ln A}{n\ln z},
\qquad 0 \leq a_{1} \leq 1/2,
\end{gathered}
\end{equation}
and
\begin{equation} \label{Theo2ab}
\begin{gathered}
\frac{1}{Mn}\ln{\mathbf P}\{S(z,M,n)\geq MA\} =
\frac{1}{Mn}\ln{\mathbf P}\{S(1/z,M,n)\geq MAz^{-n}\} \geq \\
\geq h(a_{1}) - \ln 2 - \frac{\ln(n+1)}{n}.
\end{gathered}
\end{equation}

Similarly to \eqref{Theo1a}, the relation \eqref{Theo2a} for $z \geq 1$ and
$1 \leq A \leq z^{n/2}$ can be derived.

{\bf 3. Proof of Corollary 1}. For $0 \leq z \leq 1$ we set in \eqref{Theo1}
$A = z^{n(1/2-\varepsilon)}$, $0 \leq \varepsilon \leq 1/2$. Then
$a_{0} = 1/2 -\varepsilon$, and we have
\begin{equation} \label{Cor1b}
\begin{gathered}
\frac{1}{Mn}\ln{\mathbf P}\left\{\ln\frac{S(z,M,n)}{Mz^{n/2}} \geq
-\varepsilon n\ln z\right\} \leq h(1/2-\varepsilon) -\ln 2 + \frac{\ln n}{n}
\leq -2\varepsilon^{2} + \frac{\ln n}{n},
\end{gathered}
\end{equation}
since $h(1/2-\varepsilon) \leq \ln 2 -2\varepsilon^{2}$,
$|\varepsilon| \leq 1/2$. Similarly, we have by \eqref{Theo1a} for
$0 \leq \varepsilon \leq 1/2$
\begin{equation} \label{Cor1c}
\begin{gathered}
\frac{1}{Mn}\ln{\mathbf P}\left\{\ln\frac{S(z,M,n)}{Mz^{n/2}} \leq
\varepsilon n\ln z\right\} \leq -2\varepsilon^{2} + \frac{\ln(n+1)}{n}.
\end{gathered}
\end{equation}
We set
$$
\varepsilon = \sqrt{\frac{\ln(n+1)}{n}},
$$
and then get \eqref{Cor1a} from \eqref{Cor1b}-\eqref{Cor1c}.

For $z \geq 1$ we set in \eqref{Theo2} $A = z^{n(1/2+\varepsilon)}$,
$0 \leq \varepsilon \leq 1/2$. Then $a_{1} = 1/2 -\varepsilon$, and we have
similarly to \eqref{Cor1b}
\begin{equation} \label{Cor2b}
\begin{gathered}
\frac{1}{Mn}\ln{\mathbf P}\left\{\ln\frac{S(z,M,n)}{Mz^{n/2}} \geq
\varepsilon n\ln z\right\} \leq h(1/2-\varepsilon) -\ln 2 + \frac{\ln n}{n}
\leq -2\varepsilon^{2} + \frac{\ln n}{n}.
\end{gathered}
\end{equation}
We also have by \eqref{Theo2a}
\begin{equation} \label{Cor2c}
\begin{gathered}
\frac{1}{Mn}\ln{\mathbf P}\left\{\ln\frac{S(z,M,n)}{Mz^{n/2}} \leq
\varepsilon n\ln z\right\} \leq -2\varepsilon^{2} + \frac{\ln(n+1)}{n}.
\end{gathered}
\end{equation}
We set
$$
\varepsilon = \sqrt{\frac{\ln(n+1)}{n}},
$$
and then get \eqref{Cor1a} from \eqref{Cor2b}-\eqref{Cor2c}.

\begin{center}
{\bf \S\,3. Example. Random codes}
\end{center}

In some communication systems (e.g. channels with feedback) it is useful to
follow the behavior of the true message posterior probability on time $n$.
For such purpose Corollary 1 and \eqref{Cor1a} can be applied.

Consider the binary symmetric channel ${\rm BSC}(p)$ with crossover probability
$0 < p < 1/2$, $q = 1-p$ and noiseless feedback. We consider the
case when the overall transmission time $n$ and $M=2^{Rn}$, $0 < R < 1$,
equiprobable messages $\{\theta_{1},\theta_{2},\ldots,\theta_{M_{n}}\}$ are given.
Then
\begin{equation} \label{order031}
\begin{gathered}
X_{\rm true}(n) =
\frac{\pi_{\rm true}(n)}{1-\pi_{\rm true}(n)} =
\frac{z^{d_{\rm true}(n)}}{\sum\limits_{j\neq {\rm true}}z^{d_{j}(n)}}, \qquad
z = \frac{p}{q}.
\end{gathered}
\end{equation}
It is known that
\begin{equation} \label{order032}
\begin{gathered}
\lim_{n \to \infty}\frac{1}{n}
\ln\frac{1}{{\mathbf P}\{\pi_{\rm true}(n)\neq \max\limits_{i}\pi_{i}(n)\}}
= E_{r}(R), \quad 0 \leq R \leq C(p),
\end{gathered}
\end{equation}
where
\begin{equation} \label{Erand1}
\begin{gathered}
E_{r}(R) =
\end{gathered}
\end{equation}

Without loss of generality we assume that $\theta_{\rm true} = \theta_{1}$,
and then
\begin{equation} \label{order031a}
\begin{gathered}
X_{1}(n) = \frac{\pi_{1}(n)}{1-\pi_{1}(n)} =
\frac{z^{d_{1}(n)}}{\sum\limits_{j \geq 2}z^{d_{j}(n)}} =
\frac{z^{d_{1}(n)}}{S(z,M-1,n)}, \\
S(z,M-1,n) = \sum\limits_{j=2}^{M}z^{d_{j}(n)}.
\end{gathered}
\end{equation}
By \eqref{Cor1a}, with very high probability we have $S(z,M-1,k) \sim Mz^{k/2}$
for any $k\sim n$.

We consider the probability $P(A,n)={\mathbf P}\left\{X_{1}(n) \leq A\right\}$,
$A> 0$. We have
$$
\begin{gathered}
P(A,n)={\mathbf P}\left\{X_{1}(n) \leq A\right\} =
{\mathbf P}\left\{\frac{z^{d_{1}(n)}}{S(z,M-1,n)} \leq A\right\} \sim
{\mathbf P}\left\{z^{d_{1}(n)-n/2} \leq MA\right\} = \\
= {\mathbf P}\left\{d_{1}(n) \geq \frac{n}{2}+\frac{\ln(MA)}{\ln z}\right\} \sim
q^{n}z^{na_{0}}e^{nh(a_{0})}, \\
a_{0}= \frac{1}{2}+\frac{\ln(MA)}{n\ln z}= \frac{1}{2}+\frac{Rn+\ln A}{n\ln z}.
\end{gathered}
$$
We set $A = e^{cn}$ such that
$$
\begin{gathered}
\frac{1}{n}\ln\frac{1}{P(A,n)} = -\ln q-a_{0}\ln z -h(a_{0}) = E_{r}(R), \\
a_{0} = \frac{1}{2}+\frac{R+c}{\ln z}.
\end{gathered}
$$

Note that
$$
\begin{gathered}
h(1/2-\varepsilon) = \ln 2 -2\varepsilon^{2} + O(\varepsilon^{4}) \leq
\ln 2 -2\varepsilon^{2}, \quad \varepsilon=o(1).
\end{gathered}
$$

{\bf Example 1. Random coding bounds}. Note that
\begin{equation}\label{newm022}
\begin{gathered}
{\mathbf p}(\boldsymbol{y}^{n}|\boldsymbol{x}^{n}) = q^{n}
z^{d(\boldsymbol{y}^{n},\boldsymbol{x}^{n})}.
\end{gathered}
\end{equation}

 Let $M=e^{Rn}$, $R > 0$. Then
\begin{equation} \label{order0311}
\begin{gathered}
X_{\rm true}(n) =
\frac{\pi_{\rm true}(n)}{1-\pi_{\rm true}(n)} =
\frac{z^{d_{\rm true}(n)}}{\sum\limits_{j\neq {\rm true}}z^{d_{j}(n)}}, \qquad
z = \frac{p}{q}.
\end{gathered}
\end{equation}

Обозначим через $d_{i}(k)=d(\boldsymbol{y}^{k},\boldsymbol{x}_{i}(k))$ общее
число ``отрицательных голосов'' против $\theta_{i}$ за время $[1,k]$.
Обозначим также $d_{i}=d_{i}(n)$. Тогда ($z=p/q < 1$)
\begin{equation} \label{order3}
\begin{gathered}
\pi_{i}(k) = \frac{z^{d_{i}(k)}}{\sum\limits_{j=1}^{M}z^{d_{j}(k)}} =
\frac{1}{1+\sum\limits_{j \neq i}z^{d_{j}(k)-d_{i}(k)}}, \qquad
\pi_{i}(n) = \frac{1}{1+\sum\limits_{j \neq i}z^{d_{j}(n)-d_{i}(n)}}.
\end{gathered}
\end{equation}
Заметим, что
\begin{equation} \label{order31}
\begin{gathered}
\frac{\pi_{i}(k)}{1-\pi_{i}(k)} =
\frac{z^{d_{i}(k)}}{\sum\limits_{j\neq i}z^{d_{j}(k)}} =
\frac{1}{\sum\limits_{j \neq i}z^{d_{j}(k)-d_{i}(k)}}.
\end{gathered}
\end{equation}

{\bf 1a. Upper bounds for $F(R,p)$}
The value $F(0,p)$ is known (see \cite{Bur22}). Also, it is known that
$F(R,p) \leq E_{sp}(R,p)$ for all $0 \leq R \leq C(p)$.  
Then, by \cite{CGB67}
\begin{equation} \label{upF1}
\begin{gathered}
F(R,p) \leq \left\{\begin{array}{cc}
                     E(0,p) - \lambda_{0}R, & 0 \leq R \leq R_{crit}^{f}, \\
                     E_{sp}(R,p), & R_{crit}^{f} \leq R \leq C(p),
                   \end{array}
\right.
\end{gathered}
\end{equation}
where
\begin{equation} \label{upF2}
\begin{gathered}
E_{sp}(R,p) = \rho\ln(\rho/p) + (1-\rho)\ln[(1-\rho)/q], \\
C(p) = p\ln(2p) + q\ln(2q),
\end{gathered}
\end{equation}
and in the region $0 \leq R \leq R_{crit}^{f}$ parameters $R,\rho$ and $\lambda$
are defined by relations
\begin{equation} \label{upF3}
\begin{gathered}
R = \rho\ln(2\rho) + (1-\rho)\ln[2(1-\rho)] = \ln 2-h(\rho), \qquad
p \leq \rho < 1/2, \\
1/\rho = 1+(q/p)^{1/(1+\lambda)},
\end{gathered}
\end{equation}
and the corresponding critical value $\lambda_{0}$ is defined by the equation
\begin{equation} \label{upF4}
\begin{gathered}
E(0,p) -\lambda_{0}\ln 2+ (1+\lambda_{0})\ln\left[p^{1/(1+\lambda_{0})} +
q^{1/(1+\lambda_{0})}\right] = 0.
\end{gathered}
\end{equation}

The segment $E(0,p) - \lambda_{0}R$ from \eqref{upF1}, connecting points $R=0$
and $R = R_{crit}^{f}$, is tangent to the curve $E_{sp}(R,p)$.

Note that the sphere-packing bound $E_{sp}(R,p)$ is defined by the sphere
of the  radius $d_{0}n$ such that $2^{n}/M \approx e^{nh(d_{0})}$. Then
$\ln 2 - R = h(d_{0})$ and
\begin{equation} \label{sphere1}
\begin{gathered}
\frac{1}{n}\ln P_{\rm e}(n) \geq
-E_{sp}(R,p) = \frac{1}{n}\ln{\mathbf P}\{d_{1}(n) \geq d_{0}n\}, \qquad
\ln 2 - R = h(d_{0}), \quad d_{0} = \rho.
\end{gathered}
\end{equation}

On the other hand, without loss of generality we assume that
$\theta_{\rm true} = \theta_{1}$, and then with the function $X_{1}(z,n)$
we have
\begin{equation} \label{sphere1a}
\begin{gathered}
\frac{1}{n}\ln P_{\rm e}(n) \leq
\frac{1}{n}\ln{\mathbf P}\{X_{1}(z,n) \leq 1\} \approx
\frac{1}{n}\ln{\mathbf P}\{z^{d_{1}(n)} \leq Mz^{n/2}\} = \\
= \frac{1}{n}\ln{\mathbf P}\left\{d_{1}(n) \geq
n\left(\frac{1}{2}+ \frac{R}{\ln z}\right)\right\}.
\end{gathered}
\end{equation}
The right-hand sides of \eqref{sphere1} and \eqref{sphere1a} coincide only for
the rate $R_{1} = \ln 2 -h(\rho_{1})$, where $\rho_{1}$ is defined by the
equation
\begin{equation} \label{sphere1ab}
\begin{gathered}
\left(\frac{1}{2}-\rho_{1}\right)\ln\frac{q}{p} = \ln 2 - h(\rho_{1}).
\end{gathered}
\end{equation}
For that rate $R_{1}$ we have
\begin{equation} \label{sphere1ac}
\begin{gathered}
P_{\rm e}(n) \sim P\{X_{1}(z,n) \approx 1\} = P\{\pi_{1}(n) \approx 1/2\}.
\end{gathered}
\end{equation}

In order to improve that result we use a trick. Instead of $X_{1}(z,n)$ we
consider the function $X_{1}(z_{1},n)$ with $z_{1}$ instead of $z$.
\begin{equation} \label{order0311a}
\begin{gathered}
X_{1}(z_{1},n) = \frac{\pi_{1}(z_{1},n)}{1-\pi_{1}(z_{1},n)} =
\frac{z_{1}^{d_{1}(n)}}{\sum\limits_{j \geq 2}z_{1}^{d_{j}(n)}} =
\frac{z_{1}^{d_{1}(n)}}{S(z_{1},M-1,n)} \approx
\frac{z_{1}^{d_{1}(n)}}{Mz_{1}^{n/2}}.
\end{gathered}
\end{equation}
and
\begin{equation} \label{sphere1aa}
\begin{gathered}
\frac{1}{n}\ln P_{\rm e}(n) \leq
\frac{1}{n}\ln{\mathbf P}\{X_{1}(z_{1},n) \leq 1\} \approx
\frac{1}{n}\ln{\mathbf P}\{z_{1}^{d_{1}(n)} \leq Mz_{1}^{sn/2}\} = \\
= \frac{1}{n}\ln{\mathbf P}\left\{d_{1}(n) \geq
n\left(\frac{1}{2}+ \frac{R}{\ln z_{1}}\right)\right\}.
\end{gathered}
\end{equation}

Note that \eqref{sphere1} gives the upper bound for $E(R,p)$, while
\eqref{sphere1a} gives the lower bound for $E(R,p)$. Generally, \eqref{sphere1a}
is less accurate due to the usage of level $1$. Nevertheless, if $z=z_{1}$, then
both bounds coincide when
\begin{equation} \label{sphere2a}
\begin{gathered}
d_{0} = \rho = d_{1} = \frac{1}{2}+ \frac{R}{\ln z}, \qquad
\ln 2 - R = h(\rho), \\
\ln z = -\frac{\ln 2 - h(\rho)}{1/2-\rho} \geq \ln z = \ln\frac{p}{q},
\qquad \rho \geq p.
\end{gathered}
\end{equation}

Note also that both bounds coincide for $R$, if there exists
$z_{1}=z_{1}(R) \leq 1$  such that
\begin{equation} \label{sphere2b}
\begin{gathered}
d_{0} = \rho = \frac{1}{2}+ \frac{R}{\ln z_{1}}, \qquad
\ln 2 - R = h(\rho), \\
\Rightarrow \ \ln z_{1} = -\frac{\ln 2 - h(\rho)}{1/2-\rho},
\qquad \rho \geq p.
\end{gathered}
\end{equation}


Marat V. Burnashev 

Higher School of Modern Mathematics MIPT,  1 Klimentovskiy per., Moscow, Russia \\

{\it marat.burnashev@mail.ru}

\end{document}